# Application of classical models to high resolution electric field strength falls in a hydrogen glow-discharge

V. Gonzalez-Fernandez[1,2], A. Steiger[3], K. Grützmacher[1], M. I. de la Rosa[1]

[1] *Dpto. de Física Teórica, Atómica y Óptica, Universidad de Valladolid, Paseo Belén 7, E-47011, Valladolid, Spain*

[2] *Aix-Marseille Université, CNRS, PIIM, UMR 7345, Case 321, Campus Saint-Jérôme, FR-13013, Marseille, France*

[3] *Physikalisch-Technische Bundesanstalt, Abbestr. 2-12, D-10587 Berlin, Germany*

Email: veronica.gonzalez-fernandez@univ-amu.fr

**Abstract**

This paper presents the application of two classical models to high-resolution electric field measurements carried out in a hollow cathode discharge operated in pure hydrogen plasma. The electric field determination has been done via the Stark shifting and splitting of the 2S level of hydrogen, followed by optogalvanic detection. Two classical models, Rickards' and Wroński's, are applied to these measurements with the objective of obtaining a first estimation on the discharge dynamics. The chosen models provide an idea of the ions movement, their energy and their mean free path at the cathode fall region, as well as the electric field strength behaviour depending on the discharge characteristics.

1. Introduction

The spatial and temporal distribution of the E-field is of fundamental for plasma discharge modelling [1,2]. During the last years there have been many studies focused on developing a kinetic model which explains the discharge dynamic in the cathode fall region of different kind of plasmas, different discharges geometries, pressure, etc.  (See for instance [3-7]). Obviously it is very difficult to elaborate a complete discharge model that fits to such a wide range of experimental conditions. There is still a lack of knowledge concerning several aspects of the discharges physics. Therefore high quality experimental results obtained in different discharge configurations are needed to test the models and theories, help the plasma community to improve the modelling and increase knowledge about discharge physics. This relevance of the electric field (E-field) present in low pressure plasmas has led many authors to devote their efforts into designing innovative strategies to measure it and provide reliable measures to the plasma physics community, see [8-17] and references therein. Namely, hydrogen is always interesting given their implementation in plasmas of industrial interest, fusion devices, interstellar media, etc. Moreover, atomic hydrogen is an excellent candidate for E-field measurements due to the Stark splitting of all their atomic levels is very well known [18,19].

In previous papers [20-22], we reported the determination of the strong E-field strength (ranging from about 0.5 to 4 kV/cm) in the cathode fall region of a hollow cathode discharge (HCD), operated in hydrogen (H), via the Stark splitting of the 1S - 2S transition, with a noteworthy high spatial resolution [23]. Doppler-free two-photon excitation combined with optogalvanic detection is employed to measure the E-field of the HCD in a wide range of discharge parameters. The improved reproducibility and reliability of the measurements while operating the HCD in hydrogen with tungsten cathodes was already demonstrated [21, 24]. The corresponding understanding of the degradation of HCD due the accumulation of sputtered cathode material allowed us also to apply minor corrections to the data evaluation.

In this work previous experimentally measured E-field distributions presented in [20, 21] obtained in our hydrogen HCD are compared with theoretical predictions of two classical models. In this way, we prove that these experimental results can be used to obtain useful information about the discharge characteristics, as the ions mean free path, the ion energy and the regime of the E-field strength. These parameters and their dependencies with the discharge variations (pressure and current) are analysed in this paper. This works pretends to be a first approach to the discharge modelling, therefore the chosen models present basic physical presumptions. The theories used here are described in the papers by W. D. Davis and T. A. Vanderslice [25], J. Rickards [26] and Z. Wroński [27]. All of them are focused on providing the energy of the ions bombarding the cathode in a glow discharge. Of course in the last decades much more sophisticated works have been developed, providing results of great interest [28-31], for instance. Namely, the works of Spasojević et al. [5, 32, 33] present E-field distributions models applied to plasmas similar to the one referred to in this work, but their application requires a previous knowledge of the discharge dynamics which is not available for ours at the present moment.

The present paper is structured as follows: Section 2 is devoted to explain the theorical models that will give further information about the E-field measurements and discharge dynamics. In Section 3, we present briefly the experimental arrangement where the measurements have been carried out and the E-field determination procedure. In section 4, these measurements are analyzed in the frame of the mentioned classical models, analysing all the parameters provided by each model.

2. **Theoretical aspects**

Following a chronological order, the work of Davis and Vanderslice (DV) (reference [25]) is the first one (from 1963); it is a combination of theoretical and experimental labour. It presents a simple theory with the assumption that the ions originated in the negative glow suffer symmetrical charge transfer in the cathode dark space in their way to the cathode, which are in most cases in good agreement with experiment. The charge-transfer cross sections determined are in reasonable agreement with other published data. Authors assumed in the calculations a linearly decreasing electric field across the cathode fall region.

J. Rickards [26] (in 1983) calculates the energy distribution of ions and neutrals at the cathode of a glow discharge using Monte-Carlo calculation and assuming charge exchange collisions. However he introduces different types of electric field distributions, derived from a potential proportional to $z^m$ (z is the distance from the cathode surface), according to the possible values of $m$, the obtained E-field can be linear (as in the DV model) or not. The comparison of his results with the previous model is good in the linear case.

In 1989, Z. Wroński [27], taking into account the ions motion, solved the Boltzmann equation for the ions generated in the cathode fall of an abnormal glow discharge. The solution gives an axial potential distribution, from which the E-field distribution is obtained. His results for the average ions energy are in very good agreement with the results obtained from DV model.

### 2.1 Rickards' model

In the Rickards model [26], the potential between the cathode and the nearby edge of the negative glow region increases as (as explicitly stated in [34]):

$$U(z) = U_0(1 - \frac{z}{L})^m \qquad (1)$$

z is the distance from the cathode surface, $U_0$(< 0) is the cathode fall voltage drop potential, which is approximately $U_{ac}$, this is the cathode-anode applied voltage, *L* is the length of the cathode dark space. Relative to *m* Rickard´s model considers the following cases:

- *m* = 2, corresponds to the linear E-field distribution, as in DV model [25].
- *m* = 4/3, includes the space charge effects. Model high-vacuum.
- *m* = 3/2, takes into account the mobility of ions. Model high-pressure.

The value of the *m* parameter gives not only the curvature of the E-field strength fall, but also the regime of the discharge [35].

From equation (1) the E-field distribution can be derived:

$$E(z) = U_0 \frac{m}{L}\left(1 - \frac{z}{L}\right)^{m-1} \qquad (2)$$

### 2.2 Wroński's model

Following Wroński's model [27], the axial distribution of the potential in the cathode fall is given by (as explicitly stated in [34]):

$$U(z) = -k \ln\left\{1 + \left[exp\left(-\frac{U_0}{k}\right) - 1\right] exp\left(-\frac{z}{\Lambda}\right)\right\} \qquad (3)$$

Here $\Lambda$ is the mean free path for the charge transfer, and *k* is a constant that depends on macroscopic parameters of the discharge; also it is proportional to the ion average energy in the cathode fall region, see equation (5).

From equation (3) the E-field distribution in this model results as:

$$E(z) = \frac{k}{\Lambda} \frac{\exp\left(-\frac{U_0}{k}\right) - 1}{\exp\left(\frac{z}{\Lambda}\right) + \exp\left(-\frac{U_0}{k}\right) - 1} \qquad (4)$$

Wroński model provides also the average energy of ions on their way through the cathode fall region as:

$$\bar{\varepsilon}(z) = k\left[\frac{1}{2} - \frac{1}{g(z)\sqrt{\pi}} \frac{\sqrt{\ln[g(z)]}}{\Phi(\sqrt{\ln[g(z)]})}\right] \qquad (5)$$

being:

$$g(z) = \left[\exp\left(-\frac{U_0}{k}\right) - 1\right] \exp\left(-\frac{z}{\Lambda}\right) + 1 \qquad (6)$$

and Φ(x) in (5) is the error function.
Expressions (2), (4), (5) and (6) are going to be used in this work in order to fit the experimental results.

## 3. Experimental arrangement and E-field determination

An extended explanation of the whole experimental arrangement can be found elsewhere [20], therefore only the most important details will be given here. The abnormal glow discharge is generated in a laboratory hollow cathode discharge (HCD), operated in pure hydrogen. The cathode is made of tungsten to ensure pure hydrogen plasma [24], with a length of 50 mm and 15 mm inner diameter. The cathode is placed between two stainless steel peaked anodes. Anodes and cathode have independent water cooling. This assembly provides symmetrical, long and short term stable discharges. To allow end-on spectroscopic measurements, all end pieces have axial openings. The gas enters at one anode at leaves at the opposite one, with a constant gas flow of 10 $cm^3 s^{-1}$. The plasma source can be operated in a wide range of pressures and currents: from 400 to 900 Pa and from 50 to 300 mA, leading to very different E-field strength values (from 0.5 to 4.5 kV/cm).

The measurements were performed with tuneable UV-laser radiation. It is generated in a laser spectrometer, starting in a 10 Hz injection-seeded Q-switched Nd:YAG laser (Continuum, Powerlite 9000) and a modified OPO-OPA system (Continuum, Mirage 500), followed by a sum frequency generation stage (SFG-BBO). The whole assembly provides UV-laser radiation with single longitudinal mode operation, a temporal duration of 2.5 ns and 300 MHz bandwidth. The spectra quality and regular tuning of the laser radiation is verified by an etalon of 7 GHz free spectral range and a photodiode. The standard deviation of the fringe separation (for the most usual sets of measurements) is about 160 MHz, half of the laser bandwidth.

Two opposite circularly polarized laser beams are focused in the plasma, following the selection rules of the two-photon excitation for the 1S - 2S hydrogen transition ($\Delta L = 0$). The overlapping volume is centred with respect to the cathode length and aligned parallel to the cathode surface (see Figure 1), and controlled in real time by the spatial profile analyser LaserCam-HR-UV [24]. This configuration allows performing high spatial-resolution measurements, and also Doppler-free, due to the two counter-propagating laser beams. The plasma source is mounted in two linear translators: one horizontal that controls the best alignment for the two laser beams; and one vertical that allows performing measurements at different distances from the cathode surface. The measurements start at 260 µm from the cathode surface: the minimum distance allowed by the beam divergence and the 1.53 ° crossing angle.

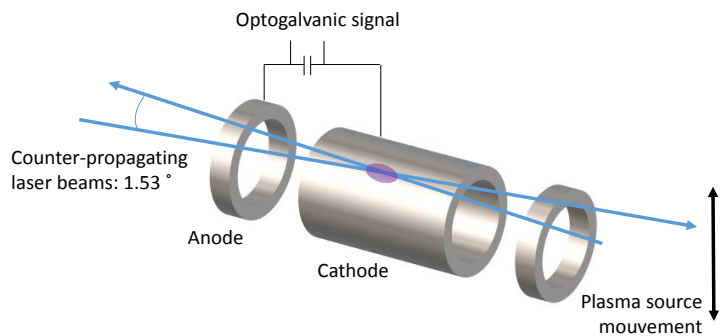

Figure 1: Scheme of the overlapping of the two laser beams, they are focalized in the upper-central part of the hollow cathode. The peaked anodes are represented as rings for a better visualization.

The simultaneous absorption of two 243 nm photons, one from each laser beam, followed by photo-ionization (due to the absorption of a third photon), induces a tiny change in the discharge impedance, that can be measured as a transient voltage drop between the cathode and one of the anodes. This technique is called optogalvanic spectroscopy. The signal largely originates from a tiny measurement volume of only 100 µm in diameter and 10 mm in length when moderate laser energy below 45 µJ is used, i.e. an irradiance of 150 MW/cm$^2$ is applied in the overlapping volume [23]. Exceeding this irradiance, the measured spectra suffer an additional broadening of the Stark components due to the depletion of the atomic ground state density and the life time reduction of the 2S level caused by photo-ionization, which, however, have no significant influence on the E-field determination when working in pure hydrogen plasmas [24].

The E-field present in the discharge causes the Stark shifting and splitting of the 2S level of hydrogen. The spectrum shows three peaks, corresponding to the $2P^{1/2}$, $2S^{1/2}$ and the $2P^{3/2}$ components. The frequency separation between the $2P^{1/2}$ and the $2P^{3/2}$ components, i.e. ($\Delta\nu=\nu(2P^{1/2}) -\nu(2P^{3/2})$), compared with theoretical calculations [36] determine the local E-field strength value. Due to the detection procedure, the signal decreases when increasing the distance from the cathode surface, and the Stark components become closer.

## 4. Application of classical models

In this section, we apply two classical models (Rickards' and Wroński's) as starting point for modelling these measurements, with the aim of obtaining a first approach to the discharge dynamics, via parameters as such as the different operation regimes, ions energy, their mean free path, etc. Of course, with the two models applied here other parameters such as maximum E-field strength and the length of the cathode fall region (already studied in previous works [21]) can be also analysed, and we also do it to confirm that the results obtained via the parabolic fit are correct. In previous works [20, 21] we reported measurements in two different cathode diameters, 10 and 15 mm. In [20], we demonstrated that keeping the same current density there was no considerable difference in the measured E-field strength. Therefore, for this work we chose 15 mm cathode, because it allows an extended cathode fall region, where more measurements could be performed. Because of that, all the measurements presented in this section were carried out in 15 mm inner diameter tungsten cathode, ensuring pure hydrogen plasma, with currents ranging from 75 to 300 mA and pressures from 400 to 900 Pa.

### 4.1. Rickards' model

As it was already mention in Section 2, this model is an evolution of the Davis-Vanderslice model [25]. This model predicts a convex electric field fall that fit very well to the experimental measurement as can be seen in Figure 2, where it is applied to E-field strength falls measured with different currents (75-300 mA) with pressures of 400, 600 and 900 Pa. As an example, the cathode fall voltage ($V_c$) and the length of the cathode fall region ($d_{cf}$) values can be directly compared with the determined in [20, 21] via the parabolic fit, obtaining a very good agreement. Besides, the E-field at the cathode surface, the maximum E-field ($E_{max}$) is very similar to the values determined via the parabolic fits, obviously a very stable parameter to characterize the discharge. This will be deeply analyzed in the following sections. The most important parameters to study here are the length of the cathode fall region ($d_{cf}$) and the *m* coefficient that gives an idea of the discharge regime.

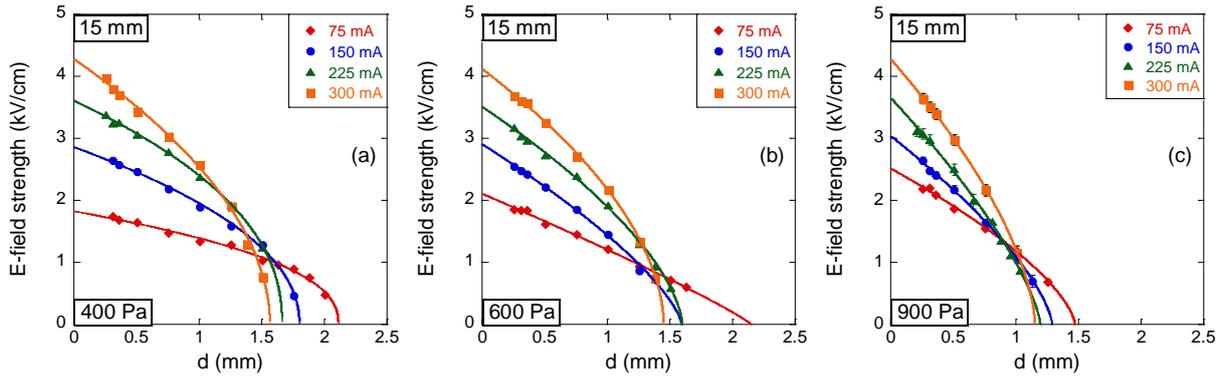

*Figure 2: E-field strength falls vs. the distance from the cathode surface, with pressures of 400 Pa (a), 600 Pa (b) and 900 Pa (c) and different current (75, 150, 225 and 300 mA), applying Rickards' model. The error bars are the same that in [20], uncertainties in the determination of the E-field.*

### *4.1.1. Length of the cathode fall region*

In this section, the $d_{cf}$ values obtained by the Rickards' model are compared with the values obtained in [21] via the parabolic fit. The result is shown in Figure 3. Due to the model curved shape, when there are no experimental data below 1 kV/cm, the last part of the fall goes down very quickly (see Figure 2). As result, the values of the length of the cathode region ($d_{cf}$) are slightly shorter than the obtained ones when a parabolic fits is applied, but they follow the same trends. As we already explained in [21], the largest values correspond to the lowest currents, and decreases when increasing current, due to the discharge structure itself. In the same way, lower pressure implies larger $d_{cf}$. The only remarkable difference between Rickards model and the parabolic fit appears in the point corresponding to 400 Pa and 225 mA. The reason is that the last experimental point is up to 1 kV/cm, and immediately, the model drops down. Ideally, more measurements would be made to complete this drop in the E-field. Unfortunately, lower values of $d_{cf}$ cannot be reached with this experimental method, because the optogalvanic signal decreases as long as we move away from the cathode surface. Lower values of the electric field, and therefore, larger values of the $d_{cf}$ can be measured with a similar spectroscopic technique, in which the 1S-3S/3D transition of hydrogen is employed; but it requires a whole different laser system (205 nm) than the one used here, so no new experimental data could be included in this work to go further. However, it is foreseen that these measures will be carried out to complement the experimental data shown so far.

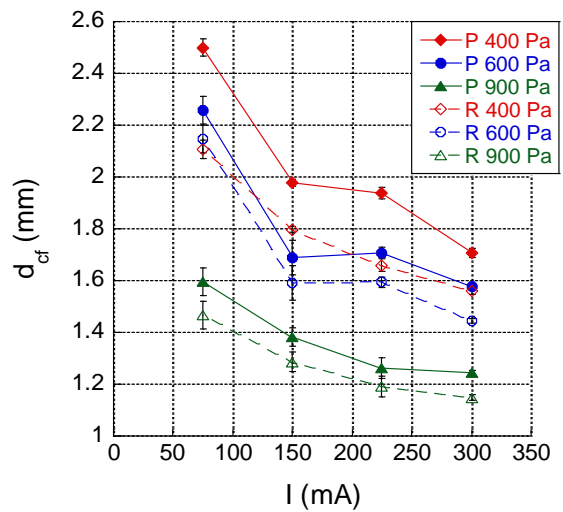

*Figure 3: Comparison of the length of the cathode fall region vs. the discharge current, obtained with the parabolic fit (solid lines) and via the Rickards' model (dashed lines). Each pressure is represented by a different colour: 400 Pa (red), 600 Pa (blue) and 900 Pa (green). The error bars are provided by the fit.*

### 4.1.2. Evolution of the coefficient m

Although simple in its presumptions, Rickards' model includes a valuable parameter regarding the discharge dynamics, the coefficient *m* that provides information about the discharge regime the curvature of the E-field strength fall [26]. This parameter changes with pressure and current, as it was explained in Section 2. A linear E-field strength fall corresponds to a *m* coefficient equal to 2, meanwhile if it approaches to 1 implies much more convex E-field strength fall.

This parameter clearly changes with the discharge conditions, see Figure 4. The evolution is represented for the three pressures and different currents values. For 400 and 900 Pa, the *m* coefficient remains almost constant, with a maximum variation of ±0.1. However, for the pressure of 600 Pa it shows a clear decreasing with increasing current. This can be verified in [20, 21], the E-field strength fall for 400 Pa is almost linear for the lowest currents, and becomes more vertical with increasing current. The different trends of the *m* coefficient for the pressures of 400 Pa and 600 Pa makes reasonable to presuppose a change in the discharge dynamics. The value of the *m* coefficient for the 600 Pa pressure decreases from 1.9 to 1.55, which would indicate that for the lowest current (75 mA) the discharge would be near the linear model (where only ion bombardment against the cathode is considered), to a high pressure model for the highest current (300 mA) where the mobility of the ions in the cathode area is taken into account until their final collision against the cathode [34, 35].

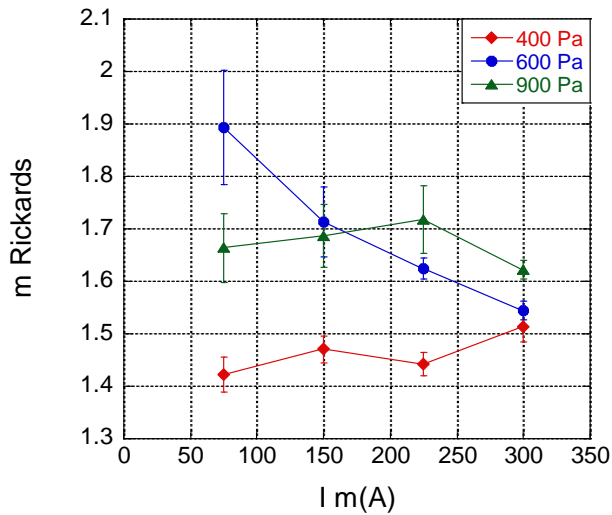

*Figure 4: Evolution of the m coefficient from the Rickards' model vs. the discharge current for three different pressures: 400 (red), 600 (blue) and 900 Pa (green). The error bars are provided by the fit.*

### 4.2. Wroński's model

We proceed now to apply Wroński's model, which presents a different trend to the employed so far. Theoretically, the E-field is practically extinguished when it reaches the negative glow. Actually, a residual E-field must remain for the maintenance of the discharge [35]. The point where this value is reached is usually accepted as the length of the cathode region ($d_{cf}$). The parabolic fit and Rickards' model predict a total drop of E-field fall, reaching zero as minimum value in the abscissa axis.

Despite Rickards or the parabolic fit, Wroński's model predicts the asymptotic fall of the E-field, which takes into account the residual E-field that remains until the discharge negative glow. In Figure 5, we show the application of Wroński's model to the E-field measurements carried out with 400, 600 and 900 Pa pressure. In general, this model takes the $d_{cf}$ value as the point the asymptotic tail has an inflection point. The maximum E-field strength remains very similar to the values obtained with the parabolic and Rickards' fits.

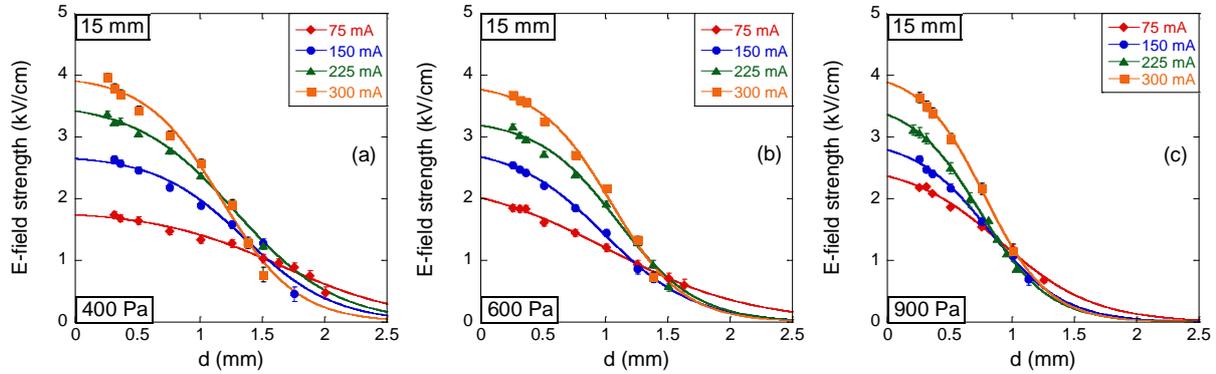

*Figure 5: E-field strengths fitted with Wroński's model vs. the distance from the cathode surface, for four different currents values (75, 150, 225 and 300 mA), with pressures of 400 Pa (a), 600 Pa (b) and 900 Pa (c). The error bars are the same that in [20], uncertainties in the determination of the E-field.*

### *4.2.1 Maximum electric field*

All the analyses performed so far indicates that the maximum E-field ($E_{max}$) appears as a stable value to characterize the discharges. To verify this presumption, in Figure 6 we present the $E_{max}$ determined via the parabolic fit (red), Rickards' models (blue) and Wroński's model (green), in 400, 600 and 900 Pa discharges. The $E_{max}$ increases linearly with the current, being Rickards values always the highest. The parabolic values lie in between of the two models, nearer to Rickards than to Wroński. The differences lie again in the shape of the models: Rickards fall is pretty much vertical at the cathode surface, meanwhile Wroński's starts a little bit more horizontal before dropping (see Figures 2 and 5). However, the maximum difference is about 4 %, therefore we conclude that the $E_{max}$ is an adequate parameter for a first discharge characterization.

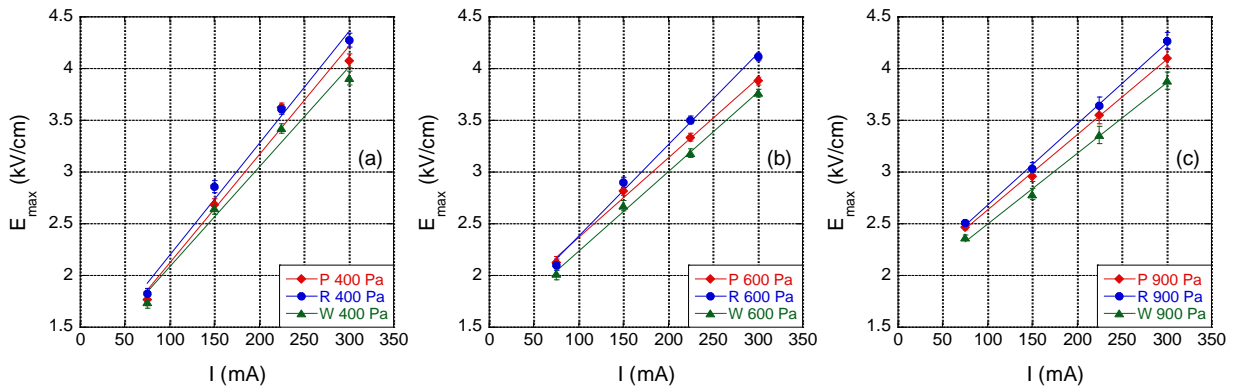

*Figure 6: Maximum E-field strength ($E_{max}$) obtained via the parabolic fit (red), Rickard model (blue) and Wroński model, vs. the discharge current, with pressures of 400 Pa (a), 600 Pa (b) and 900 Pa (c). The error bars are provided by the different fits.*

*4.2.2. Ions energy*

To obtain a general understanding of the discharge, it is important to know the behaviour of the different ions ($H^+$, $H_2^+$, $H_3^+$) in the cathode region. Unfortunately, setting global values is very complicated because ion dynamics are specific to each type of discharge. However, different authors have obtained very interesting results about ion dynamics and dominant ionic species in each discharge region, either via experimental measurements or simulations [7, 34, 37-40]. In our approximation to the discharge dynamics, Wroński model offers access to parameters concerning the particles dynamics; this makes the model so interesting for a first approach.

Equation 5 allows obtaining an estimation of the mean ion energy in the cathode region. It is important to take into account that Wroński model does not distinguish between the different ionic species that coexist there. Therefore, the energies calculated here cannot be assigned to a single ion type; they have to be considered as the average of the net charge present in that area. The energies calculated are shown in Figure 7, where the three pressures of 400, 600 and 900 Pa are analysed. As can be seen, energy decreases as long as the ions go away from the cathode surface, being this decrease more abrupt with the current. The fact that the energy is higher for the lower pressure is linked to the fact that the ions could accelerate more when they have a larger cathode area. The minimum energy is almost the same for all currents and pressures, about 10 eV.

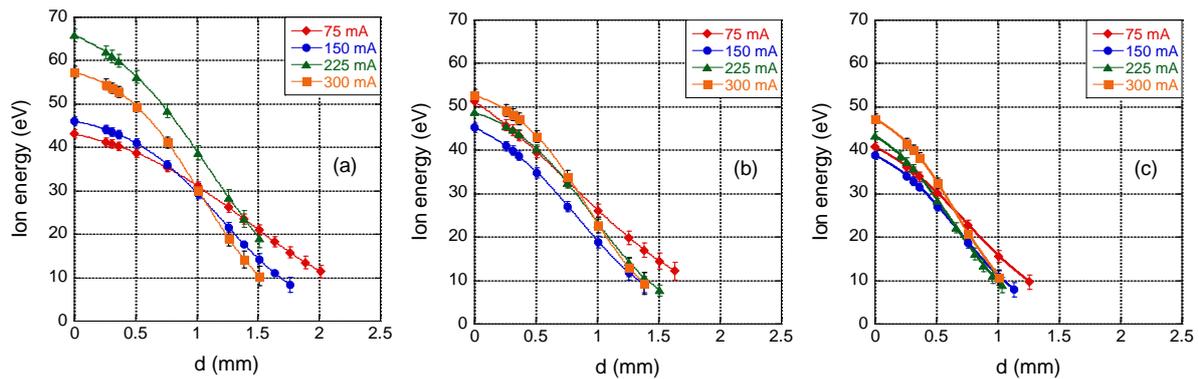

*Figure 7: Ion energy in the cathode fall region vs. the distance from the cathode surface, discharge with with pressures of 400 Pa (a), 600 Pa (b) and 900 Pa (c) and four different currents (75, 15, 225 and 300 mA) obtained by the Wroński model. The error bars are provided by the fit.*

It is more realistic to estimate the ions energy at the cathode surface, due to a smaller disparity of ionic species is present there [34]. The estimated energies are shown in Table 1. The maximum values oscillate between 35 and 70 eV, depending on the discharge conditions. As special feature, we can highlight that the maximum current does not lead to a maximum energy; it can be one of the medium currents. In the case where the maximum current leads to a maximum energy, the medium currents suffer a little decrease. These fluctuations correspond to the variations of $E_{max}$ that were already observed and explained in a previous article (see Figure 4(b) of [21]): the largest $E_{max}$, and therefore the maximum ion energy, does not increase linearly with pressure, it suffers a slight decrease in the 600 Pa pressure. The results are in concordance with other authors, and near ideal gas conditions [7, 34].

| I (mA) | Energy 400 Pa (eV) | Energy 600 Pa (eV) | Energy 900 Pa (eV) |
|---|---|---|---|
| 75 | 43.1 ± 1.0 | 51.3 ± 1.2 | 40.8 ± 1.0 |
| 150 | 46.2 ± 1.0 | 45.3 ± 1.04 | 38.9 ± 0.9 |
| 225 | 65.9 ± 1.5 | 48.8 ± 1.05 | 43.4 ± 1.0 |
| 300 | 57.5 ± 1.2 | 52.8 ± 1.4 | 47.3 ± 1.1 |

*Table 1: Ions energy at the cathode surface for different pressures (400, 600 and 900 Pa) and several currents (75, 150, 225 and 300 mA), obtained by the Wroński model.*

### *4.2.3. Mean free path of ions*

Wroński model also provides a value of the ions mean free path in the cathode region. It is quite difficult to give a physical meaning to this parameter, and it has to be considered as a first reference for further studies, as it is also indicated in [34]. The mean free path is shown in Figure 8, where the values decrease with increasing pressure and current. The maximum value is reached with the lowest current and pressure, and they decrease with increasing pressure and current. This is coherent with the compression suffered by the cathode fall region when pressure and current increase, that has been already demonstrated in [21]. The point corresponding 400 Pa and 225 mA is slightly higher, differing from the general trend. As we already explained in Section 4.1.1., this is due to the fact that the last experimental point of this series of measurements lies around 1 kV/cm (see also Figure 3(c) of [21]). As we have already mentioned, the accuracy of these models relies heavily on having experimental measurements that cover the complete E-field fall. In this specific case, the measurements end earlier, and the model begins to create the asymptotic curve at the moment when no more experimental points are available. There is also another point to remark: at 75 mA the mean free path is larger for 600 Pa that for 400 Pa. This seems not very intuitive, but we need to remember that all the evidences point to a change in the discharge regime between these two pressures.

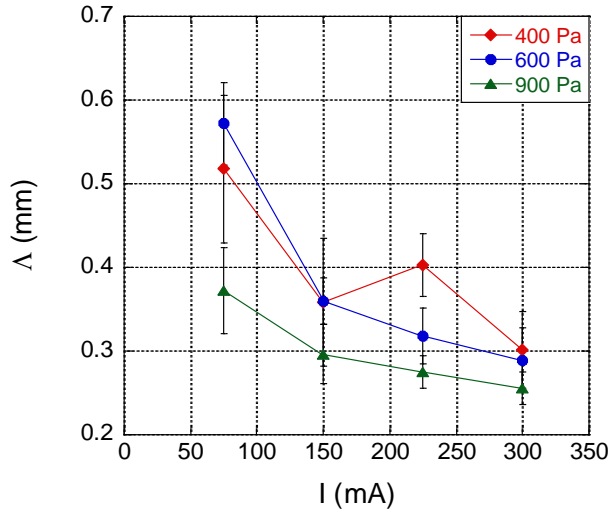

*Figure 8: Mean free path of the ions in the cathode fall region vs. the discharge current, for three different pressures (400, 600 and 900 Pa). The error bars are provided by the fit.*

### 5. Summary

In this paper we present a basic characterization of a hydrogen glow-discharge carried out in a 15 mm tungsten cathode via the application of two classical models: Rickards and Wroński models. Both models provide different discharge parameters that give a first idea of the plasma behaviour.

Rickards model allows the determination of the *m* coefficient that modulates the E-field strength fall. It gives an idea of a change in the discharge regime when the pressure increases from 400 to 600 Pa. From this model the length of the cathode fall region $d_{cf}$ is also obtained and compared with the value provided by the parabolic employed until now, confirming the same trends. The values coming from the Rickards model are systematically slightly shorter than the parabolic ones, due to the more abrupt end of the model when arriving to the end of the cathode region.

From Wroński model parameters such as the ions mean free path and its average energy in the cathode fall region are obtained. It is important to keep in mind that these two parameters represent an average value of all ionic species that can be found in the cathode region. However, the results obtained in this work are in good agreement with other authors, and near ideal gas conditions. These values show how the ions start from the cathode with a high amount of energy that is lost progressively in their path to the centre of the discharge, due to collisions with other particles. The ion mean free path, in general, decreases with pressure and current, as does the length of the cathode zone.

On the other hand, we show that the maximum E-field strength $E_{max}$ is a suitable parameter for an initial discharge characterization regardless of the fit used.

The results achieved in this article verify the conclusions in our previously published works, since all the trends observed here are coherent with the results already shown. The results also confirm the conclusions of previous works by providing more information about the state of the discharge.

Finally, we would like to emphasize again that this work is a starting point when speaking about discharge dynamics. The two classical models have been successfully applied to experimental data as a first step towards a deeper understanding of hydrogen glow discharges. Further understanding of the discharge dynamics can be obtained, not so much through pressures and intermediate currents (already explored in previous work of the group, such as [22]), but through measurements in deuterium. As it is an isotope with the same characteristics as hydrogen, only mass changes, it can provide very interesting information about the ions movement at the cathode fall region and their energy. Deuterium will be employed as buffer gas in the next measurement campaign.

**Acknowledgments**

The authors thank DGICYT (Ministerio de Economía y Competitividad) for the project ENE2012-35902, FEDER funds and the grant BES-2013-063248 given to V. González-Fernández. The authors thank A. Martín and S. González for the technological support, J. L. Nieto for the informatics help, E. M. Domingo for the administrative work.